# Room-Temperature Electron-Hole Liquid in Monolayer $MoS_2$

Yiling Yu[1#], Alexander W. Bataller[2#], Robert Younts[2], Yifei Yu[1], Guoqing Li[1], Alexander A. Puretzky[3], David B. Geohegan[3], Kenan Gundogdu[2*], Linyou Cao[1,2,4*]

[1] Department of Materials Science and Engineering, North Carolina State University, Raleigh, North Carolina, 27695, United States; [2] Department of Physics, North Carolina State University, Raleigh, North Carolina, 27695, United States; [3] Center for Nanophase Materials Sciences, Oak Ridge National Laboratory, Oak Ridge, Tennessee 37831, United States; [4] Department of Electrical and Computer Engineering, North Carolina State University, Raleigh, North Carolina, 27695, United States

\# These authors contributed equally

\* Correspondence should be addressed to: lcao2@ncsu.edu and kgundog@ncsu.edu

**Abstract**: Excitons in semiconductors are usually non-interacting and behave like an ideal gas, but may condense to a strongly-correlated liquid-like state, *i.e.* electron-hole liquid (EHL), at high density and appropriate temperature. EHL is a macroscopic quantum state with exotic properties and represents the ultimate attainable charge excitation density in steady states. It bears great promise for a variety of fields such as ultrahigh-power photonics and quantum science and technology. However, the condensation of gas-like excitons to EHL has often been restricted to cryogenic temperatures, which significantly limits the prospect of EHL for use in practical applications. Herein we demonstrate the formation of EHL at room temperature in monolayer $MoS_2$ by taking advantage of the monolayer's extraordinarily strong exciton binding energy. This work demonstrates the potential for the liquid-like state of charge excitations to be a useful platform for the studies of macroscopic quantum phenomena and the development of optoelectronic devices.

Many-body interactions of elementary particles consists of the foundation of forming complex matter. The interaction of atoms or molecules gives rise to the classical physical states of matter, *i.e.* solid, liquid, and gas. Similarly, the strong correlation of quantum particles or quasi-particles like electrons, phonons, excitons, and spins can lead to the formation of exotic states of matter that exhibit macroscopic quantum phenomena. Notable examples include superconductivity,[1] Bose-Einstein condensation,[2] Mott insulators,[3] and electron-hole liquid.[4] These macroscopic quantum states provide extraordinary electrical, optical, and thermal functionalities, and bear great promise for applications in a wide range of fields from electronics to photonics, sensing, and energy. However, unlike the classical physical states, which may be stable at room temperature or above, macroscopic quantum states are typically only stable at cryogenic temperatures.[1-4] This feature has significantly limited the prospect of these states for use in practical applications.

Here we demonstrate the formation of a macroscopic quantum state of strongly-correlated charge excitations known as electron-hole liquid (EHL) at room temperature in monolayer $MoS_2$. In semiconductors, the pairing of electrons and holes *via* Coulomb interaction can result in an exciton.[5] At low densities, excitons are non-interacting and behave like an ideal gas, which are hence referred to as "free excitons" (FEs).[5] But at high densities, the correlation and exchange energies of excitons can lead to a phase transition from FE to EHL, in which a large number of quantum degenerate electrons and holes are bound by collective interaction rather than through pair-wise Coulomb attraction.[4] This phase transition is a quantum analog to the liquidation of a classical gas. It is a first-order transition and only occurs at appropriate charge density and temperature, analog to the requirement of the classical gas liquidation for appropriate pressure and temperature. EHL is a Fermi liquid exhibiting many features similar to a classical liquid, such as incompressibility, surface tension, and short-range order. Much like the difference between classical liquids and gases, EHL shows properties dramatically different from FEs [6-10] and bears great promise for the studies of many-body quantum phenomena and the development of functional devices like broadband high-power lasers and quantum computing. The observation of room-temperature EHL in our experiment is facilitated by the strong quantum confinement and reduced dielectric screening in monolayer $MoS_2$ due to its atomically thin dimension. A direct result is a strong exciton binding energy $E_b$. According to prior studies in bulk semiconductors, the thermal stability of EHL is governed by an empirical relationship of $k_B T_c \sim 0.1 E_b$, where $k_B$ is the Boltzmann constant and $T_c$ is the critical temperature beyond which EHL boils into FEs, much like the boiling point of a classical liquid. [4, 11-14] Previous studies have also shown that the EHL in conventional direct or indirect bandgap semiconductors is only stable at cryogenic temperatures due to small values of $E_b$ (*e.g.*, ~ 15 meV in silicon[15]).[8, 9, 13, 16] In contrast, the high $E_b$ (around 0.42 eV[17]) in monolayer $MoS_2$ is expected to enable stable EHL at room temperature and even above ($T_c \sim$ 500K). Another key reason for the successful observation of room-temperature EHL in our experiment is the utilization of suspended monolayer $MoS_2$, which eliminates the effect of substrates on exciton dynamics. Substrates are able to strongly affect the dynamics of excitons in monolayer $MoS_2$.[18-20] For example, the lifetime of excitons in suspended $MoS_2$ can be an order of magnitude longer than substrate-supported $MoS_2$.[18]

**Results and Discussion**

We start by examining the photoluminescence (PL) of suspended monolayer MoS$_2$, as PL spectral analysis has historically been the key diagnostic for evaluating the presence of EHL.[8, 11-14, 21, 22] Fig. 1a shows the spatial-spectral images of the PL collected from suspended MoS$_2$ monolayers illuminated by a 532 nm laser with different power densities. The monolayer was grown on sapphire substrates using a chemical vapor deposition process, and then transferred onto quartz substrates that are pre-patterned with 6 μm diameter cavities (see Fig. S1 and Materials and Methods in Supporting Information).[23, 24] The incident laser has a Gaussian profile and was focused to a spot size of ~35 μm, which is much larger than the suspended monolayer. In particular, we examine the PL spectra obtained from an area of ~ 2 μm$^2$ located at the center of the suspended monolayer (see Materials and Methods in Supporting Information). From the resulting PL spectra (Fig. 1b), we extract the integrated intensity, peak position, and peak width (full width at half maximum, FWHM) and plot them as a function of the power density in Fig. 1c-d. It is worth noting that the experimental results are very reproducible. We have confirmed no quality degradation in the materials during the measurement, and repeatedly observed similar results with multiple measurements at different samples (see Fig. S2-S4 and §S1 of Supporting Information).

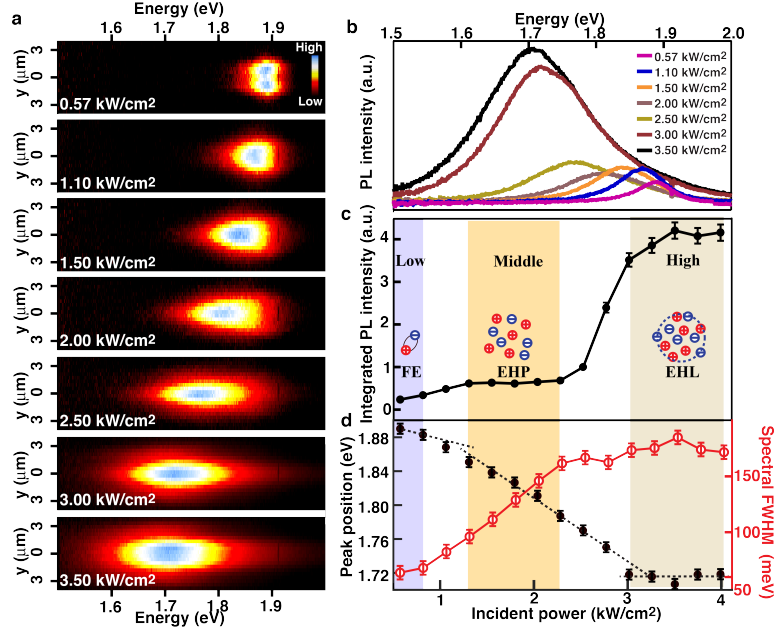

**Fig. 1. Power dependence in the PL of suspended monolayer MoS$_2$.** (**a**) two-dimensional image for the spatial-spectral profile of PL under different incident power densities. The incident power density is labelled as shown. (**b**) PL spectra extracted from the center of the illuminating area. (**c**) Dependence of the PL intensity on incident power. (**d**) Peak position (black) and FWHM (red) of the PL as a function of incident power. The colored areas indicate low, middle, and high incident power regions, in which excitons exist in format of free excitons (FE), electron hole plasma (EHP) and electron-hole liquid (EHL), respectively.

We find that the PL strongly depends on the incident power density. Based on the PL's spectral features, we roughly divide the entire power range studied into three regimes: low, middle, and high photoexcitation that are connected by two transition stages as indicated in Fig.1c-d. As discussed following, the distinct PL features are due to the evolution of the physical state of excitons with the incident power, which are free excitons (FEs), electron-hole plasma (EHP), and electron-hole liquid (EHL) at the low, middle, and high power regimes, respectively. In the low power regime (< 0.80 kW/cm$^2$), the PL exhibits a sharp peak around 1.89 eV with a FWHM of ~50 meV (Fig. 1b and Fig. 2a). It is the characteristic emission of monolayer MoS$_2$ from the *A* exciton at the K and K′-points of the Brillouin zone. The PL intensity increases with the incident power in an approximately linear way, and the peak position and spectral width show very mild variation with increasing power, which is consistent with the previous extensive studies for the *A* exciton emission of monolayer $MoS_2$.[18, 20, 25] In contrast, the PL intensity is roughly power-independent in the middle power regime (1.30-2.30 kW/cm$^2$). In this regime, the PL also undergoes substantial spectral redshifting and broadening with increasing power. To better illustrate the difference in PL, Fig. 2a plots typical PL spectra obtained in the low and middle power regimes. Compared to the low power emission, the PL in the middle power regime shows absence of the sharp exciton resonance, and experiences a large spectral redshifting (~130 meV) and broadening (~200 meV FWHM). We can exclude laser heating as the major cause for the redshift and broadening, because the laser heating effect on the PL spectra is much smaller (~40 meV redshift and ~35 meV broadening, see temperature analysis in Fig. S5-S7 and §S2 of Supporting Information).

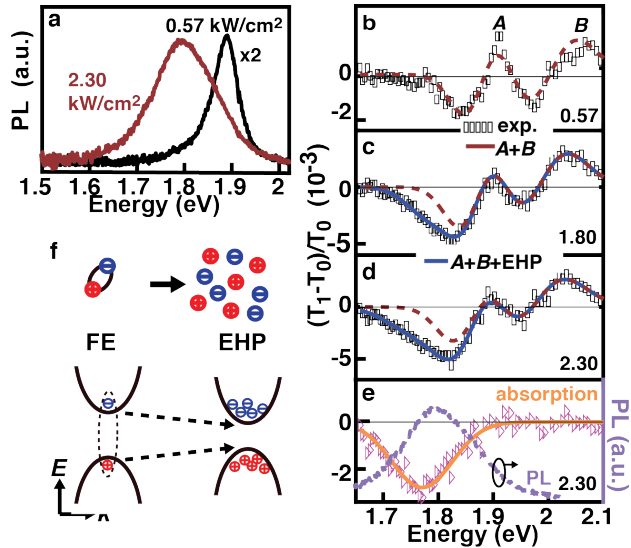

**Fig. 2. Comparison of the PL and absorption collected at the low and middle power regimes.** (a) Comparison of typical PL spectra of suspended monolayer $MoS_2$ in the low and middle power regimes. Two powers are used as examples, 0.57 kW/cm$^2$ and 2.30 kW/cm$^2$. The intensity of the PL collected at 0.57 kW/cm$^2$ is scaled by a constant of 2 for convenience of comparison. Measured and fitted differential absorption spectra of suspended monolayer $MoS_2$ at **(b)** 0.57 kW/cm$^2$, **(c)** 1.80 kW/cm$^2$ and **(d)** 2.30 kW/cm$^2$. The differential absorption is indicated by the transmission difference in monolayer $MoS_2$ with ($T_1$) and without ($T_0$) the incident laser as ($T_1$-$T_0$)/$T_0$. The black rectangles are experimental results, the red dashed curves are the fitting for the differential absorption contributed by the *A* and *B* excitons, and the blue solid curves are the fitting for the differential absorption contributed by the *A* and *B* excitons plus the EHP. **(e)**

Absorption and PL (dashed purple curve) of EHP under incidence of 2.30 kW/cm$^2$. The scattered triangles indicate the measured absorption of EHP, obtained by subtracting the dashed red line from the scattered rectangles in (c). The yellow solid line is the fitting result. **(f)** Schematic illustration for the transition from free excitons (FEs) to electron-hole plasma (EHP) as well as the associated bandgap renormalization.

We correlate the PL in the middle power regime to the emission of electron-hole plasma (EHP). To illustrate this notion, we measured differential absorption spectra of the suspended monolayer MoS$_2$ in the presence of laser photoexcitation. The differential absorption is defined as the difference between the optical absorption of the monolayer with and without illumination of the incident laser. We studied the differential absorption with transmission measurement, because previous studies of ultrathin 2D materials on transparent substrates show that a moderate change in transmission or reflectance is mainly determined by the absorption of the materials.[26] More specifically, we measured the transmission of a weak broadband white light through the monolayer with or without the presence of the incident laser. In the low power regime, we observe a decrease in the absorption of both *A* and *B* excitons as indicated by positive signal in the measured differential transmission, and also observe an increase in the absorption for the light with energy lower than the excitons as indicted by negative signal in the differential transmission (Fig. 2b). They are due to photobleaching and photo-induced absorption, respectively, and consistent with the results of previous transient absorption measurement at MoS$_2$.[27, 28] In the middle power regime, we observe the emergence of a broad absorption peak (negative signal) in an energy region substantially lower than the optical bandgap of MoS$_2$ (Fig. 2c-d). For instance, the absorption peak for the 2.30 kW/cm$^2$ photoexcitation is located at 1.76 eV with a FWHM of around 200 meV. The spectral position and width of this absorption reasonably match the spectral features of the PL obtained at the same incident power (Fig. 2e), which strongly suggests a common origin. We exclude charged exciton, multi-excitonic complexes (such as bi-excitons or tri-excitons) or the formation of defects as the reason for the observed sub-bandgap absorption/emission for the following reasons: 1) the PL intensity in the middle power regime is power-independent (in contrast, the intensity of charged exciton and multi-excitonic emission is expected to increase super-linearly with incident power[29-31]), and 2) the FE spectral shape and intensity at low powers remains unchanging after the monolayer has been subjected to higher laser fluences (Fig. S3).

The two major features we observed in the middle power regime, namely, the absence of the sharp exciton resonance and the emergence of broad, low energy absorption/emission features, have also been observed in previous semiconductor studies with a high density of electron-hole pairs[11, 26, 32, 33] and ascribed to the formation of a gas-like state of partially or fully ionized excitons, *i.e.* an electron-hole plasma (EHP). The transition of excitons to EHP is referred as an excitonic Mott transition and results from the screening of the Coulomb attraction between electrons and holes due to the presence of a high carrier density (Fig. 2f). This screening reduces the exciton binding energy and lead to the dissociation (also referred as ionization) of excitons.[11, 32, 34] The formation of EHP could account for the experimental trends observed in the middle power regime. The ionization of excitons and the reduction of exciton binding energy due to electrostatic screening results in the observed absence of the excitonic resonance. The presence of ionized excitons may induce a bandgap renormalization as illustrated in Fig. 2f, which gives rise to a redshift in the emission/absorption peaks. Additionally, unlike the emission of FEs, which is sharp because only the electrons and holes at the band edge are involved, the emission of EHP involves all the electrons in the conduction band and the holes in the valence band and hence feature a relatively broader spectral width.[32, 34] Both the bandgap renormalization and spectral width of EHP are associated with the density of charge carriers.[32] With the incident power increasing, the density of ionized excitons (electrons and holes) increase, leading to larger bandgap renormalization and broader spectral width in emission/absorption. This accounts for the observed continuous red-shifting and broadening in PL with increasing photoexcitation. Of the most interest is the PL at the high power regime (> 3.00 kW/cm$^2$). The PL intensity shows a sharp threshold-like increase by 4-5 folds during the transition from the middle to high power regimes (2.50-3.00 kW/cm$^2$) in which the incident powers only increases by 18% (Fig. 1b-c). Additionally, the PL in the high power regime exhibits a constant peak position and spectral width regardless the incident power. This is in stark contrast to the continuous redshifting and broadening of the PL in the middle power regime. To better understand the PL in the high power regime, we examine the spatial profile of the PL. Fig. 3 show the PL spatial images (Fig. 3a) and the spatial distribution of the PL intensity across the suspended monolayer that are extracted from the images (Fig. 3b-c). The PL intensity at the center initially increases with the incident power. It reaches a maximum value and does not change with the incident power at the high power regime. While the power increase at the high power regime does not increase the PL intensity, it causes the region showing the maximum PL intensity (referred as the maximum region) to expand. Additionally, when the incident powers reaches 3.50 kW/cm$^2$, the maximum region evolves from a flat-top to a ring-like distribution in which the center shows lower PL intensity than the edge. The PL at the ring center is also narrower and blueshifted compared to the PL at the ring edge (Fig. S8). With increase in the incident power, the PL at the ring center turns to be even weaker, narrower, and more blueshifted, while the PL at the edge shows no change. This indicates that the charge carrier density at the ring center is lower than the edge and decreases with increasing photoexcitation. Again, we exclude changes in the composition and crystalline structure of the monolayer (such as defect creation) during the measurement, as we have confirmed the preservation of the low-power PL and Raman features of the monolayer after the measurement in the high power regime (see Fig. S3-S4).

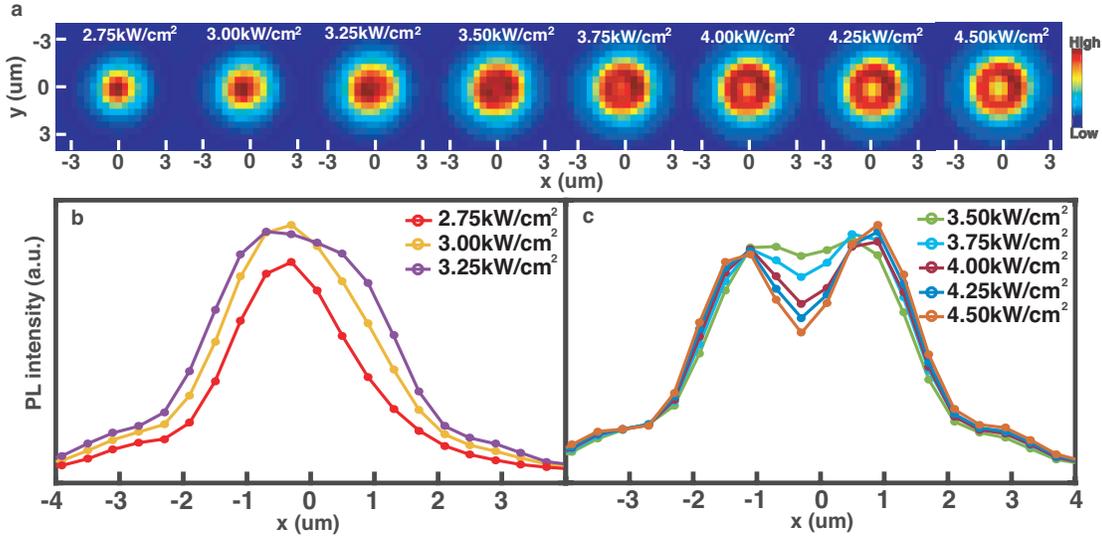

**Fig. 3. Spatial profile of the PL in the high power regime.** (**a**) Spatial images of the luminescence area at different incident power densities. The incident power density is labelled as shown. (**b-c**) The spatial distribution of PL intensity across the suspended monolayer at different incident power densities.

We correlate the PL in the high power regime to the emission of EHL because all the spectral and spatial features, which includes the substantial enhancement in intensity, the constant peak position and spectral width, the spatial expansion of the maximum region, and the formation of a luminescence ring, are consistent with the main features of EHL previously reported.[6-9, 13, 34-38] First of all, it is known that EHP may experience a first-order phase transition into EHL at a critical charge density and under appropriate temperature. Compared to EHP, EHL features stronger correlation energy between electrons and holes, and thereby may have stronger PL efficiency.[34, 39] This is consistent with the sharp threshold-like improvement of PL efficiency observed in the transition from the middle to high power regimes. Second, one key difference between EHL and EHP lies in the incompressibility of EHL. The charge density of EHP increases with incident power, and thereby the EHP emission continuously redshifts and broadens as observed in experiments. In contrast, EHL is incompressible with a constant charge density regardless the incident power.[4, 11, 34] This is consistent with the constant peak position, spectral width, and intensity of the PL observed in the high power regime, as all these features are dictated by the density of charge carriers.[34, 39] As a result of the incompressibility, a further increase in incident power cannot increase the charge density of EHL, but can increase in the volume of the EHL.[4,34, 40] This matches our observation that the further increase in photoexcitation beyond 3.00 kW/cm$^2$ does not lead to either an increase in the maximum PL intensity or further spectral redshifting and broadening, but just causes expansion of the EHL area. Third, the formation of the luminescence ring can be correlated to another important feature of EHL, high mobility. EHL is highly mobile due to the reduction of friction with the crystal lattice by the Fermi degenerate charge carriers.[4, 39, 41] It may move through crystal lattices with the presence of non-uniform deformation, strain, electric/magnetic fields, or intense excitation.[4, 39, 41] Usually, the density of photo-excited charges is determined by three dynamic processes, charge generation (absorption of incident power), diffusion, and recombination, but these processes cannot enable a decrease of charge density with increasing photoexcitation as observed in the luminescence ring. The only possible way to enable the luminescence ring and its dynamic evolution with incident power is an outward delocalization of the charge generated at the center. This delocalization must overwhelm the concentration gradient-induced diffusion and can move charges from low density to high density regions. It implies a high mobility of the excited charges as expected for EHL. Although more studies are necessary to better understand the physics underlying the ring formation, we believe the heavy incidence might give rise to a non-uniform distribution of temperature, strain, and/or bandgap, which subsequently drive the delocalization of EHL to form the ring. To further support the formation of EHL, we examine the charge density and related physical features, including chemical potential and PL spectral profile. Charge density is the single most important parameter to dictate the formation of EHL in our experiment. Similar to the liquidation of a classical gas, which is driven by chemical potential, the formation of EHL is driven by a chemical potential that is associated with charge density $n$ and temperature $T$. The temperature is required to be below the critical temperature $T_c$, while the charge density must reach a certain critical value to enable a minimal chemical potential. The extraordinarily high $E_B$ of monolayer MoS$_2$ gives rise to a high value of $T_c$ (~500K), and this makes the temperature requirement be readily satisfied in our experiment (see temperature analysis in §S2 of Supporting Information). Therefore, the remaining criterion for THE EHL formation in our experiment is achieving the critical charge density. Intuitively, much like a classical liquid in which neighboring molecules are in touch, the critical charge density of EHL is expected to enable physical overlap of the quantum wave functions of neighboring charges, which gives rise to an interparticle distance in the order of the exciton diameter $2a$ as $n \sim (2a)^{-2}$.[21, 39, 41] Additionally, the critical charge density is expected to enable a chemical potential lower than that of FEs, the difference of which serves as the necessary thermodynamic driving force for the condensation of FEs to EHL.[21, 39, 41] The critical charge density is also expected to dictate the spectra width of the PL, as the PL involves the contribution from all the electrons and holes.[34] In the following discussion, we estimate the charge density for monolayer MoS$_2$ in the high power regime from experimental measurements, and find the estimated value matches all the above-mentioned expectations for the critical charge density of EHL.

We estimate the charge density for monolayer MoS$_2$ in the high power regime using two different methods, which results in similar values. Without losing generality, we use 3.00 kW/cm$^2$ as an example for the discussion. We first estimate the charge density $n$ based on the charge lifetime $\tau$, absorption efficiency $\alpha$, and incident power $I$ as $n = \alpha\tau I$. We have previously demonstrated that the charge lifetime in suspended monolayer MoS$_2$ is around 100 ns when the incidence power is high enough to enable the formation of EHL ( Note: the EHL was referred as electron-hole plasma condensate and dense electron hole plasma in the previous work due to insufficient evidence that time), [42] which is two orders of magnitude longer than that of free excitons.[18] Similar results showing much longer lifetime of EHL than free excitons has also been reported at Ge previously.[39] The high density of charges might effectively suppress defect-assisted recombination and exciton-exciton annihilation, which are the two major recombination mechanism in monolayer MoS$_2$,[18, 43, 44] by filling in the defect states and screening Coulomb interactions, just like what was reported for Ge.[39] The absorption efficiency of suspended monolayer MoS$_2$ for the incident laser is 0.062.[17] The charge density is then estimated at $4.2 \times 10^{13}$ cm$^{-2}$. Additionally, we may estimate the charge density $n$ from the change of bandgap induced by the presence of free charges $\Delta E_{BG}$, which is usually referred as bandgap renormalization. Previous studies in quantum wells have demonstrated that $\Delta E_{BG}$ in quasi-2D systems is correlated with the charge density $n$ as well as the exciton radius $a$ and binding energy $E_b$ of excitons as $\Delta E_{BG} = -3.1(na^2)^{1/3}E_b$.[45] $\Delta E_{BG}$ is the difference between the original and renormalized electronic bandgaps, the former being equal to the sum of optical bandgap (1.89 eV) and exciton binding energy $E_b$ (0.42 eV) and the latter being the low energy band bottom taken from the measured PL spectra as indicated in Fig. 4a. Using the result in Fig. 4a and the exciton radius and binding energy obtained from our previous experiments ($a$ = 0.65 nm, and $E_b$ = 0.42 eV),[17] we estimate the charge density in monolayer MoS$_2$ at 3.00 kW/cm$^2$ to be $3.4 \times 10^{13}$ cm$^{-2}$ (see §S3 of Supporting Information), which is consistent with the one estimated from the charge lifetime.

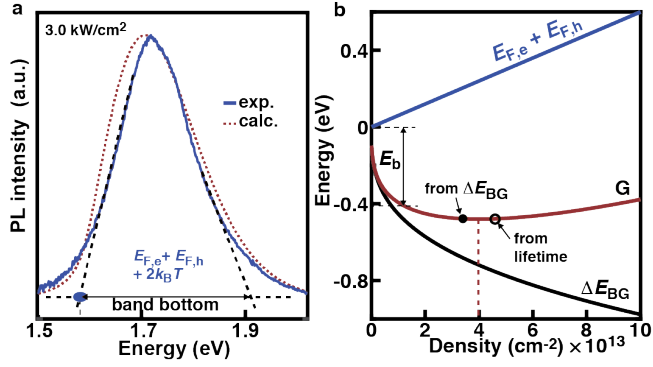

**Fig. 4. Analysis for the EHL in monolayer MoS$_2$.** (**a**) Measured and simulated PL spectra at the incident power of 3.00 kW/cm$^2$. The cross points of the dashed lines indicate the band bottoms of the spectra. The energy difference between the low energy edge and high energy edge of the spectra is equal to the sum of the Fermi energies and the thermal energies as labeled. (**b**) Free energy (red curve) of ionized excitons in monolayer MoS$_2$ as a function of the charge density. Also plotted are the Fermi energies of the electron and holes (blue curve) and the bandgap renormalization (black curve). The bottom of the original conduction band is set to be the reference point (zero energy), and the vertical black line indicates the binding energy of free excitons $E_b$. The charge densities estimated from experimental measurements are labeled in the curve, from the band renormalization (black solid dot) and from the charge lifetime (black circle). The vertical red dashed line indicates the minimal free energy.

The critical charge density estimated from our experiments also agree with the theoretical calculation based on free energy minimization of EHL. The free energy of a many-body electron-hole system $G$ is related with the Fermi energies of electrons $E_{F,e}$ and holes $E_{F,h}$ and the bandgap renormalization $\Delta E_{BG}$ as $G = E_{F,e} + E_{F,h} + \Delta E_{BG}$.[4, 46] Each of these terms have analytic or empirical relationships with the charge density $n$ (see §S3 of Supporting Information).[45, 47] Fig. 4b shows the calculated free energy $G$ as a function of the charge density $n$ in monolayer MoS$_2$. EHL is reported to form at the minimum free energy.[21, 39, 41] In our calculation, the minimum free energy has a charge density of $4.0\times10^{13}$ cm$^{-2}$, consistent with the value estimated from the experimental measurement. To better illustrate the physics, we ignore the effect of temperature and focus on the ground state free energy in the theoretical calculation. This approximation is justified due to the very small thermal energy relative to the transition energy. Furthermore, the Fermi energy of monolayer MoS$_2$ calculated without considering the effect of temperature shows a reasonable agreement with the experimental measurements at room-temperature.[47] The bandgap renormalization is also insensitive to temperature[45]. The ground state free energy is equal to the binding energy.[21, 39, 41] The result in Fig. 4b indicates that the EHL has an energy of 0.48 eV per electron-hole pair. This agrees with the intuitive expectation that the minimum free energy of EHL is lower than the FE state, as necessitated by the thermodynamic driving force for the phase transition from FEs to EHL.[21, 39, 41]

The charge density estimated from the experimental measurement also matches the length scale expected for the critical charge density of EHL and is consistent with the observed PL spectral profile. The estimated charge density ($3.4\times10^{13}$ cm$^{-2}$) is related to the exciton radius $a$ by $n = (2.6a)^{-2}$, reasonably consistent with the expected density-distance correlation for EHL $n \sim (2a)^{-2}$. Using our measured charge density, we calculate the EHL PL spectra using a luminescence model that involves all the occupied electronic states in the conduction band and the hole states in the valence band (see §S3 of Supporting Information).[4, 34] Fig. 4a shows reasonable agreement between the calculated (dashed line) and measured (solid line) PL emission collected at 3.00 kW/cm$^2$. The emission width (defined by the difference between the low energy edge and the high energy edge) of the spectra is equal to the Fermi energies of electrons and holes plus the thermal energy, which is consistent with previous EHL reports.[4, 34] This further confirms the validity of the estimated charge density and supports the formation of EHL.

**Conclusion**

In summary, we have demonstrated the condensation of non-interacting, gas-like excitons to a strongly-correlated, liquid-like state, *i.e.* electron-hole liquid (EHL) at room-temperature in monolayer MoS$_2$. The formation of the room-temperature EHL benefits from the extraordinary strong exciton binding energy of monolayer $MoS_2$ and the suspension of the monolayer from surrounding substrates. The EHL is realized by injecting high densities of excitons with heavy incidence. With increasing photoexcitation, excitons first experience a Mott transition to become an ionized state known as EHP, and then condenses into EHL with the charge density reaching a critical value. EHL is a Fermi liquid consisting of degenerate electrons and holes bound by collective interactions, and represents the ultimate attainable density of charge excitations in steady states. The formation of EHL is evidenced by PL measurements that exhibit key characteristics of previously reported EHL. These include a sharp threshold-like enhancement in PL efficiency, incompressibility in charge density (constant PL intensity/peak position/spectral width and expansion of the area with the maximum PL intensity), high mobility, and a value of the critical charge density that may give rise to a minimal free energy and an interparticle distance in the order of the exciton radius $a$ as $n \sim (2a)^{-2}$. Similar room-temperature exciton condensation is expected to exist in other 2D materials with strong exciton binding energies such as $WS_2$, $MoSe_2$, and $WSe_2$. Room-temperature EHL could provide a model system to study the fundament physics of many-body interactions and macroscopic quantum states. Additionally, it may enable the development of devices, such as ultra-high-power super-broadband lasers or LEDs and quantum computing devices. During the paper reviewing process, we have become aware of similar work of room-temperature EHL in $MoTe_2$ that was published in Nature Photonics doi.org/10.1038/s41566-019-0349-y.

**Methods/Experimental**

**Substrate preparation, and monolayer transfer** The monolayer (obtained from 2Dlayer) was grown on sapphire substrates using a well-established chemical vapor deposition process[24] and then was transferred onto the quartz substrates with pre-patterned holes using a surface-energy-assisted transfer approach.[23] The holes were defined using standard photolithography and dry etching processes. In a typical transfer process, 9 g of polystyrene (PS, MW=280 kg/mol) was dissolved in 100 mL of toluene. The PS solution was spin-coated (3000 rpm for 60 s) on the as-grown monolayer. This was followed with 80-90° C baking for 1 hr. A water droplet was then dropped on top of the monolayer. Water molecules may penetrate under the monolayer to lift off the PS-monolayer assembly. The polymer-monolayer assembly was then transferred to the quartz substrate using tweezers. The sample was baked at 80°C for 1 hour and then another 30 min at 150° C. The PS was removed by rinsing with toluene.

**Laser excitation and imaged photoluminescence (PL)** PL measurements were performed using a 532 nm TEM$_{00}$ diode laser focused to 35 μm. PL was collected by a 50× objective and imaged through a 570 nm longpass filter and onto the entrance slit of a spectrometer. The spectrometer output was recorded with a PIXIS CCD cooled to -80° C. Spatial PL was recorded by keeping the entrance slit fully open and moving the grating to the 0$^{th}$ order mode. Spectrum was recorded by closing the entrance slit to 2 μm in the imaged dimension and moving the grating to the 1$^{st}$ order diffraction mode. All the PL measurements were done under vacuum (30 mTorr).

**Differential absorption spectrum (DAS) and fitting** DAS is a quasi-steady-state measurement where a modulated pump laser excites the system, while a broadband laser probes the changes in absorption from photo-generated populations. The DAS was measured using a non-collinear pump-probe spectroscopy technique. Suspended monolayer MoS$_2$ was pumped with a modulated CW TEM$_{00}$ 532 nm laser. A broadband white light continuum probe (580 – 750 nm) was generated using a femtosecond Ti:Sapphire laser (rep. rate = 250 kHz, λ = 830 nm) focused through a quartz window. The probe was focused to < 2 μm with a 50× objective and transmitted through the hole center of the suspended monolayer, and then spectrally resolved using a monochromator and Si photodiode. The change in transmission as a function of wavelength was measured with a lock-in amplifier and recorded. All DAS measurements were done in vacuum (30 mTorr). DAS were fit using a reduced $\chi^2$ minimization to the following equation

$$DAS(E) = A_{EHP}e^{-(E-E_{EHP})^2/\sigma_{EHP}^2} + A_{EX_a}(E - E_{EX_a})e^{-(E-E_{EX_a})^2/\sigma_{EX_a}^2} \\ + A_{EX_b}(E - E_{EX_b})e^{-(E-E_{EX_b})^2/\sigma_{EX_b}^2},$$

where the EHP is represented by a Gaussian distribution and each A and B FE state is represented as a derivate of a Gaussian distribution with respect to energy. The possible contribution from trions is ignored for simplicity, because trions have been shown negligible in suspended monolayers as the doping level and net charges are minimal. In the fitting, we also keep the amplitude ratio, frequency difference, and spectral width ratio of the *A* and *B* excitons to be pretty much constant, as this is what expected in experiments due to the close correlation between the *A* and *B* excitons. Parameter error is estimated by varying the parameter of interest until the reduced $\chi^2$ varies by 1, yielding the 1-σ boundaries.

**Acknowledgments**
This work was supported by the National Science Foundation under the grant of ECCS-1508856 and DMR 1709934. The authors acknowledge the use of the Analytical Instrumentation Facility (AIF) at North Carolina State University, which is supported by the State of North Carolina and the National Science Foundation. Part of the Raman and PL work was conducted at the Center for Nanophase Materials Sciences, which is sponsored at Oak Ridge National Laboratory by the Scientific User Facilities Division, Office of Basic Energy Sciences, U.S. Department of Energy.

**Competing financial interests**
The authors declare no competing financial interests.

**Supporting Information Available:** < Reproducibility of the power-dependent PL measurement; Evaluation of temperature and its effect during the measurement; Theoretical calculation for the EHP and EHL. > This material is available free of charge *via* the Internet at
http://pubs.acs.org

TOC

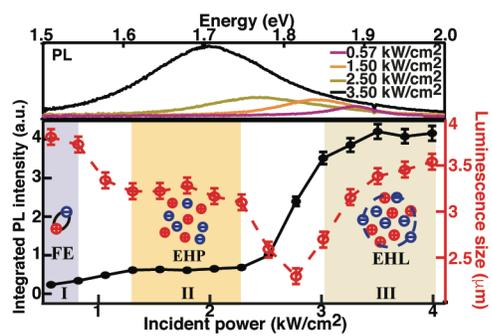